# Aerosol parameters for night sky brightness modelling estimated from daytime sky images

M. Kocifaj,[1,2]★ F. Kundracik[2] and J. Barentine[3]

[1]*ICA, Slovak Academy of Sciences, Dúbravská cesta 9, 845 03 Bratislava, Slovakia*
[2]*Department of Experimental Physics, Faculty of Mathematics, Physics, and Informatics, Comenius University, Mlynská dolina, 842 48 Bratislava, Slovakia*
[3]*Dark Sky Consulting, LLC, Tucson, AZ 85730-1317, USA*

**ABSTRACT**

Atmospheric turbidity is one of the key factors influencing the propagation of artificial light into the environment during cloudless nights. High aerosol loading can reduce the visibility of astronomical objects, and thus information on atmospheric pollution is critical for the prediction of the night sky brightness (NSB) distribution. In particular, the aerosol optical depth (AOD) and asymmetry parameter ($g$) are among the most important aerosol properties influencing the NSB amplitudes. However, these two parameters are rarely available at astronomical sites. Here, we develop a method for AOD and $g$ retrievals from clear-sky radiometry carried out around sunset or sunrise, shortly before or after night-time observation is intended. The method allows for reducing the number of unknowns needed in the processing and interpretation of night sky radiances, and thus provides an efficient tool for gathering input data to present skyglow simulators. The practice of collecting information about aerosols in this way could become a routine part of astronomical observations, much like observing standard stars to obtain extinction coefficients. If the procedure were conducted around sunset and the data were quickly reduced, it could offer an on-the-spot estimate of the NSB for the night ahead. The error analysis is performed using the theoretical model, while taking into account experimental errors of radiance readings. The capability of the method is demonstrated in a field experiment conducted under cloudless conditions.

## 1 INTRODUCTION

Night sky brightness (NSB) resulting from outdoor lighting is among the most critical factors that hinder astronomical observations in many localities. The negative impacts of light pollution can be mitigated by implementing novel technological solutions for outdoor lighting (Schroer & Hölker 2017), but the utility of this strategy is ultimately limited. This is because the propagation of artificial light into the ambient environment is efficiently modulated by the momentary state of the atmosphere. Under clear-sky conditions, atmospheric turbidity is by far the most important parameter governing the temporal variability of the NSB (Kocifaj & Barentine 2021). The essential elements needed to estimate the sphere of influence of ground-based light sources include, at least, the aerosol optical depth (AOD) and the aerosol asymmetry parameter ($g$) (Aubé 2015; Wallner & Kocifaj 2023; Bará, Rigueiro & Lima 2019). When combined with the optical depths of atmospheric gases, AOD predetermines the optical attenuation of light beams traversing the cloud-free atmosphere. Elevated values of AOD are also known to cause an increase of the diffuse background (Van de Hulst 1980). This is because the energy removed from the initial flux of photons is spatially redistributed, giving rise to the diffuse light. Here the parameter $g$ plays an important role and acts as a form factor for the NSB. The scattered photons concentrate around the direction of beam propagation when $g$ approaches unity, while side-scattering becomes more pronounced for low values of $g$. The theoretical values of $g$ range from –1 for backscattering to +1 for light scattered exclusively in the forward direction. However, negative values of $g$ are not common for the Earth's atmosphere. Furthermore, $g = 0$ for isotropic scattering is rare. Most typically, $g \gtrsim 0.4$, as discussed below equation (9).

Predicting and evaluating the effects of artificial light sources on skyglow in astronomical sites requires information on $g$ and AOD. However, sensors providing such optical data are generally sparsely deployed across large territories (Tegen et al. 2019). For instance, only a few AERONET (AErosol RObotic NETwork) stations operate in central Europe (Gonzi et al. 2007), and only a single AERONET site is located in Slovakia (in Poprad-Gánovce). Continuous active (Lewis et al. 2013) and passive (Zhang et al. 2023) night-time optical sensing of the atmosphere is still rare, so the corresponding AOD and $g$ values are typically unavailable and are thus inferred indirectly based on information on the prevailing types of air pollution sources and their distribution around measuring sites. Nevertheless, in periods of high atmospheric pressure, during stable cloudless synoptic conditions persisting over hours to days, it is convenient to perform optical measurements shortly before sunset or shortly after sunrise. Both experiments can provide supplementary information about aerosols, which are in turn applicable to NSB modelling during night-time hours. A reasonable agreement between night-time AOD values approximated based on straddling daytime-averaged AOD values and high-resolution aerosol sensing has also been documented in a satellite observation study (McHardy et al. 2015), in which night-time AOD was interpolated from late-afternoon and early-morning AOD data. Ground-based methods of daytime sky imaging seem to be even more advantageous because they are well established, easy to conduct at arbitrary sites using portable devices, and capable of providing the required information on AOD and *g*.

★ E-mail: kocifaj@savba.sk

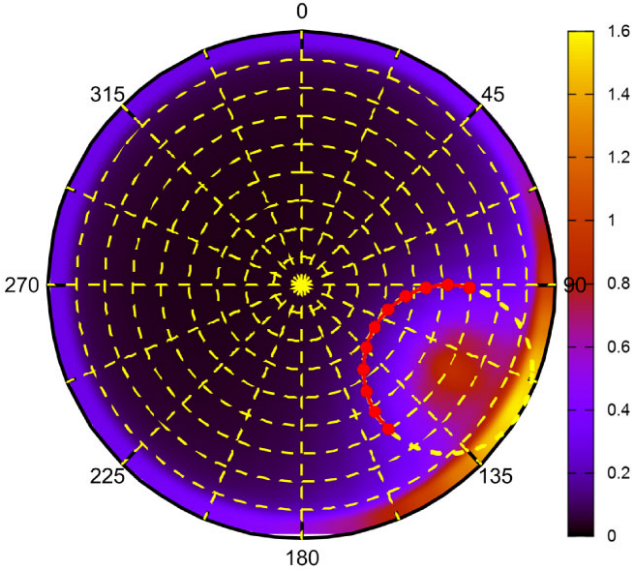

**Figure 1.** A false colour image of the total clear-sky daytime sky radiance in W m$^{-2}$ nm$^{-1}$ sr$^{-1}$ calculated using the UNISKY model (Kocifaj 2015) and determined as a superposition of single- and double-scattered light (wavelength $\lambda = 550$ nm, solar zenith angle $z_0 = 60°$, solar azimuth angle $= 120°$, AOD $= 0.3$, ROD $= 0.1$, ground albedo $= 0.2$, aerosol asymmetry parameter $g = 0.7$). The azimuth is measured in a clockwise direction (with north at the top). The zenith and horizon are in the centre and at the edge of the plot, respectively. Radiance data are plotted on a linear scale. The circular zone encloses the sky elements preferred for experiments. Red circles are for sky elements at a fixed angular distance from the solar disc.

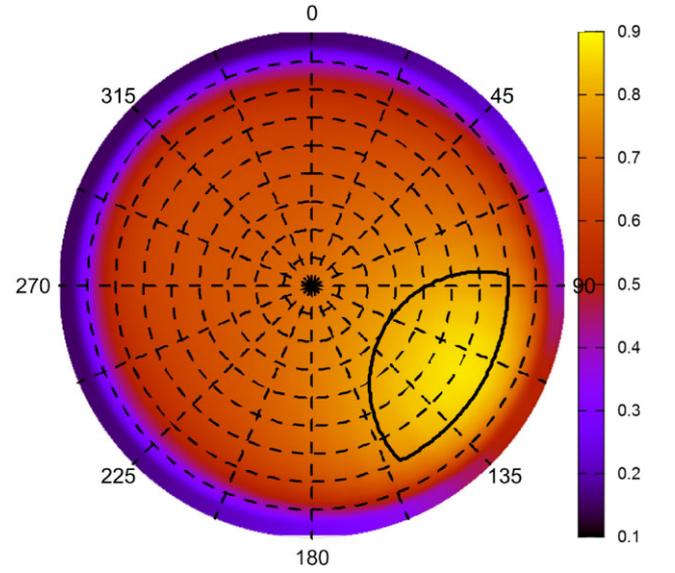

**Figure 2.** The ratio of single- to total-scattering radiance computed for the same input parameters as in Fig. 1. The black solid line encloses the sky area with the dominant contribution to the total radiance from single scattering to the total radiance.

Here we present an easily realizable method to extract AOD and *g* from clear-sky radiance distribution data taken shortly before sunset or after sunrise.

## 2 THE MODEL

Radiative transfer modelling of the Earth's atmosphere can be difficult owing to the multiple scattering of light, which introduces a theoretical complexity to the model and makes the numerical simulations CPU-intensive. During clear-sky conditions, the 3D radiative transfer equation can be reduced to a 1D problem for a horizontally homogeneous atmosphere (Tapimo et al. 2021). Multiple scattering typically shapes the clear daytime sky radiance at large angular distances from the Sun; however, these effects are much less pronounced for sky elements near the Sun. Here, the major contribution to sky radiance is attributable to single scattering (Nakajima et al. 2020), for which a number of analytical solutions exist for a plane-parallel atmosphere.

For low Sun positions (such as that shown in Fig. 1), we used the unified model of radiance patterns (Kocifaj 2015) to compute both single- and double-scattering radiances of the daytime sky and found that single scattering dominates in a circular area with radius of up to 30°–40° around the Sun, except for sky elements near the horizon

(see Fig. 2). In the region of the figure enclosed by the black solid line, the sky radiance *L* is approximated as follows:

$$L(z) = \pi S \tilde{\omega} \frac{P(\theta)}{4\pi} \frac{\cos z_0}{\cos z - \cos z_0} \left\{ e^{-\tau \sec z} - e^{-\tau \sec z_0} \right\}, \quad (1)$$

where $\pi S$ is the flux density of solar radiation at the top of the atmosphere; $P(\theta)$ is the atmospheric scattering phase function at an angular distance $\theta$ from the solar disc; $\tau$ is the atmospheric optical depth, which measures the attenuation of light within the atmospheric column; $\tilde{\omega}$ is the average atmospheric single-scattering albedo; and $z_0$ and $z$ are the solar and observational zenith angles, respectively. For most daytime sky scanners, $\tau$ can be determined as the sum of the AOD and Rayleigh optical depth (ROD); that is, $\tau = AOD + ROD$. The impact from gaseous and water vapour absorption bands is negligible until a radiometric measurement is intentionally made within a specific absorption peak. In the solar almucantar (i.e. for $z = z_0$), equation (1) reduces to

$$L(z_0) = \pi S \tilde{\omega} \tau \frac{P(\theta)}{4\pi} \frac{e^{-\tau \sec z_0}}{\cos z_0}. \quad (2)$$

For more details, see also the derivations in section 5.1 of Lenoble (1985); the second term of the right-hand side of equation (8) in Duan, Min & Stamnes (2010); and equations (7) and (8) in Russkova, Sviridenkov & Zhuravleva (2016).

## 3 EXPERIMENTAL DEMONSTRATION OF THE METHODOLOGY

Looking at the form of equations (1)–(2), it is convenient to perform daytime sky radiance measurements along the solar vertical plane (the straight line from the Sun towards the zenith) and at a constant angular distance from the solar disc (circular arc), as shown in Fig. 3. An example for discrete measurement directions is indicated in the figure. The ratio of sky radiances taken along the circular arc at different zenith angles $z_i$ allows for isolating the contribution from the atmospheric optical depth, because the scattering phase function is cancelled out when computing $R(z_1, z_2) = L(z_1)/L(z_2)$. Let us consider that $z_1 = z$ is for a sky element in the solar vertical plane, while $z_2 = z_0$ is for the solar almucantar. In such a case, the above ratio becomes

$$R(z, z_0) = \frac{1}{\tau} \frac{\cos^2 z_0}{\cos z - \cos z_0} \left\{ e^{-\tau(\sec z - \sec z_0)} - 1 \right\}, \quad (3)$$

which is a form independent of the scattering phase function and thus independent of $g$. The best-fitting $\tau$ is then determined by matching the experimental and theoretical ratios $R(z, z_0)$.

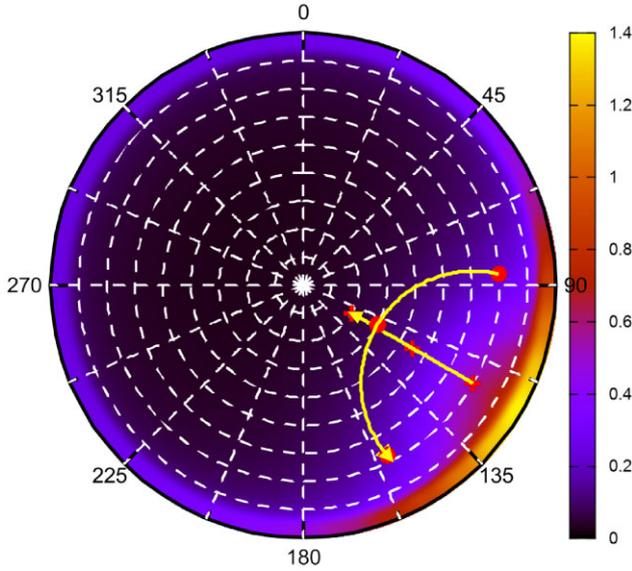

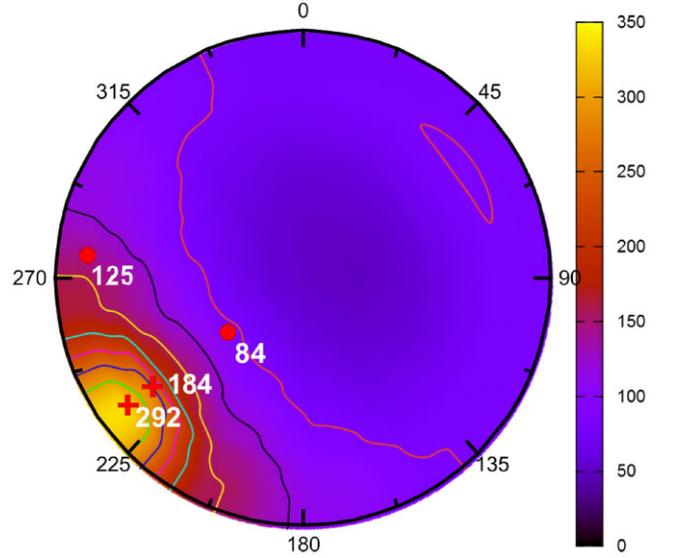

**Figure 3.** As Fig. (1), but for $z_0 = 70°$. Two measurements are necessary to obtain AOD and $g$. First, day sky-brightness sensing is required for a set of discrete (at least two) points along the circular arc. An unknown contribution from the scattering phase function is eliminated this way, because all points are located at the same angular distance from the solar disc. Therefore, $\tau$ and AOD can be obtained by minimizing differences between the measured and computed ratios of $L(z)/L(z_0)$. Second, the measurement along the solar vertical plane should be carried out to obtain $g$. The minimum of two points along yellow straight line have to be selected (avoiding the direction to the solar disc because direct sunbeams are excluded).

**Figure 4.** All-sky daytime radiance for $\lambda = 450$ nm measured in Bratislava on 2017 October 16. Unlike in Figs 1–3, the maximum measured zenith angle (at the edge of the plot) is now $z = 80°$. Observations near the horizon have been excluded because different blocking obstacles were located in multiple azimuthal directions. Sky elements displayed as filled red circles are selected for the purpose of using equation (3) aiming to determine $\tau$. Red crosses represent the directions of the radiances needed to retrieve $g$. The symbols used here and in Fig. 3 are identical. Radiance data are not in absolute units. Calibration to the standard unit system (W m$^{-2}$ nm$^{-1}$ sr$^{-1}$) is not necessary because the coefficient of proportionality is cancelled out when dividing the radiances of two different sky elements.

If $z_1$ and $z_2$ are chosen along the solar vertical, the ratio of day sky radiances is

$$R(z_1, z_2) = \frac{P(\theta_1)}{P(\theta_2)} \frac{(\cos z_2 - \cos z_0)}{(\cos z_1 - \cos z_0)} \left( \frac{e^{-\tau \sec z_1} - e^{-\tau \sec z_0}}{e^{-\tau \sec z_2} - e^{-\tau \sec z_0}} \right), \quad (4)$$

where $\theta_1 = z_0 - z_1$ and $\theta_2 = z_0 - z_2$ are the respective scattering angles. The atmospheric scattering phase function $P(\theta)$ is obtained as a weighted combination of the Rayleigh term $P_R(\theta)$ and the aerosol component $P_A(\theta)$. The weighting factors are linearly proportional to AOD and ROD, as shown below:

$$P(\theta) = \frac{AOD \times P_A(\theta) + ROD \times P_R(\theta)}{AOD + ROD}. \quad (5)$$

Here

$$P_R(\theta) = \frac{3}{4} \left( 1 + \cos^2 \theta \right) \quad (6)$$

(Fan et al. 2021) and

$$ROD \approx 0.00879 \lambda^{-4.09} \quad (7)$$

(Teillet 1990).

Modelling light scattering from atmospheric aerosols is not a trivial task because of the wide range of particle sizes, shapes, compositions and spatial orientations observed in nature (Mishchenko

2014). Although the scattering phase function of a monodisperse or uniform single-morphology particle population tends to show specific scattering signatures with a set of peaks and minima, a smooth form of $P_A(\theta)$ is typical for random discrete media such as atmospheric aerosols. Among various analytical approximations for $P_A(\theta)$, the Henyey–Greenstein function ($P_{HG}$) is frequently applied in aerosol optical studies (Winkler 2022; Kómar, Wallner & Kocifaj 2022; Sabater et al. 2021). Here

$$P_{HG}(\theta) = \frac{1 - g^2}{(1 + g^2 - 2g \cos \theta)^{3/2}}. \quad (8)$$

Substituting $P_A(\theta)$ for $P_{HG}(\theta)$ and inserting equations (5)–(8) into equation (4), the parameter $g$ can be found by minimizing the difference between the measured and theoretical values of $R(z_1, z_2)$; that is, by comparing the left-hand and right-hand sides of equation (4).

In Fig. 4 we present the results of a daytime experiment conducted in the late afternoon hours of 2017 October 16–17, using a portable sky scanner that operates linearly over 5–6 orders of magnitude for a range of narrowband filters (Kocifaj, Kómar & Kundracik 2018). The solar zenith angle ($z_0$) and azimuth angle were 73° and 234°, respectively. Likewise in Fig. 3, filled circles are for different $z$, but for a fixed angular distance from the solar disc ($\theta = 40°$, see Table 1). Red crosses are for zenith angles $z_1 = 60°$ and $z_2 = 70°$, both measured along the solar vertical plane. The figure depicts spectral radiances for the wavelength of $\lambda = 450$ nm, displayed in relative units. Absolute calibration is not required because the coefficient of proportionality cancels out when determining the ratios $R(z, z_0)$ and $R(z_1, z_2)$.

The best-fitting atmospheric optical thickness of 0.55 was found when plotting $R(z, z_0)$ as a function of $\tau$; see the left-hand plot in

$$\frac{P(\theta_1)}{P(\theta_2)} = R(z_1, z_2) \frac{(\cos z_1 - \cos z_0)}{(\cos z_2 - \cos z_0)} \left( \frac{e^{-\tau \sec z_2} - e^{-\tau \sec z_0}}{e^{-\tau \sec z_1} - e^{-\tau \sec z_0}} \right) \quad (9)$$

is isolated from equation (4). For the given Sun position and sky elements selected, the method is found to have sufficient sensitivity to asymmetry parameters $g \gtrsim 0.4$, which in fact covers the typical range of clear-sky states observed worldwide (e.g. Kinne et al. 2013; Hatzianastassiou et al. 2007; Kinne 2019). Nevertheless, it is suggested that the method should be tested for sensitivity with respect to the parameter $g$ each time it is used. The positions of the sky elements should be modified if the value of $g$ approaches the flat part of dependence displayed in Fig. 5. The solution is stable and unique owing to the monotonic form of the two functions depicted in Fig. 5. ROD is 0.23 at $\lambda = 450$ nm, so AOD is $\approx 0.32$ (0.55–0.23). AOD and $g$ obtained from the two nearest AERONET stations, located in Vienna (AOD $\approx 0.25$, $g \approx 0.72$) and Brno (AOD $\approx 0.22$, $g \approx 0.72$), are of similar magnitude, although the distances to Bratislava range from 50 to 120 km. The synoptic situation was stable, with a ridge of high pressure extending over tens to hundreds of kilometres. For this reason, we expect that AOD was more or less stable over a large territory. Vienna has almost no heavy industry, and most pollution sources have excellent filtering systems installed. Unlike in Bratislava, the major source of particles in Vienna is probably traffic. The average concentration of PM2.5 in Bratislava is roughly 62 $\mu$g m$^{-3}$, while the corresponding value for Vienna is 47 $\mu$g m$^{-3}$ (Neuberger et al. 2004). Therefore, we would expect AOD in Bratislava to be generally higher than that in Vienna in an optically stable atmosphere.

**Table 1.** The radiance values taken in Bratislava in the afternoon hours (15:30–16:30) of 2017 October 16–17. The data readouts are in relative units and correspond to those displayed in Fig. 4.

|  | Zenith angle (°) | Azimuth angle (°) | Radiance (value read) |
| --- | --- | --- | --- |
| Red cross no. 1 | 70.0 | 234.0 | 292.0 |
| Red cross no. 2 | 60.0 | 234.0 | 184.0 |
| Red circle no. 1 | 70.0 | 276.0 | 125.0 |
| Red circle no. 2 | 30.0 | 234.0 | 83.7 |

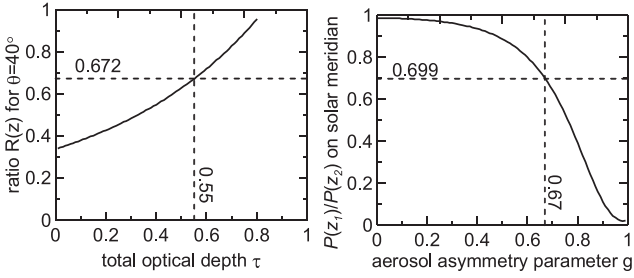

**Figure 5.** The theoretical values of $\tau$ and $g$ that best satisfy the experimental conditions represented in Fig. 4. Left-hand plot: the ratio $R(z, z_0)$ plotted as a function of $\tau$ (the angular distance from the solar disc is $\theta = 40°$). The best-fitting model value of the aerosol optical depth for the measured ratio of 0.672 is $\tau = 0.55$. Right-hand plot: the ratio of scattering phase functions via $g$. The horizontal dashed line is for observations made at $z_1 = 60°$ and $z_2 = 70°$. This corresponds to $g = 0.67$.

Fig. 5. The data reading at the intersection point has the experimentally determined value of $R(z, z_0) = 84/125 = 0.672$. Analogously, the approximate value of the aerosol asymmetry parameter, 0.67, was extracted from right-hand plot of Fig. 5, in which $P(z_1)/P(z_2)$ is shown as a function of $g$. The ratio of scattering phase functions

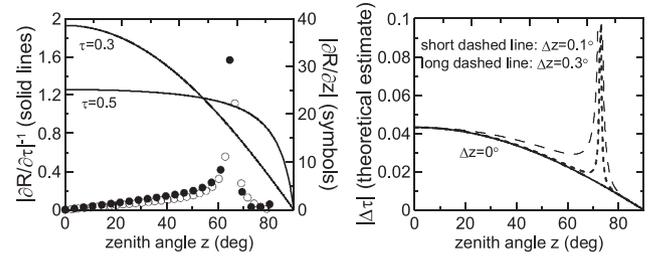

**Figure 6.** Left-hand plot: modelled $|\partial R/\partial \tau|^{-1}$ and $|\partial R/\partial z|$, both shown as functions of zenith angle. Solid lines are for $|\partial R/\partial \tau|^{-1}$ computed for two optical depths, $\tau = 0.3$ and $\tau = 0.5$ (see labels on left-hand side). Circles are for $|\partial R/\partial z|$ (open circles are for $\tau = 0.3$; solid circles are for $\tau = 0.5$). Right-hand plot: theoretical estimate for atmospheric optical depth computed from equation (11) assuming a relative error in radiance ($\Delta L)/L(z_0) \approx 0.1$ (see equation 15). The error in the position of a sky element is considered to vary from 0° to 0.3°. The value of $\Delta z = 0°$ is used because the angular resolution error is usually negligible for professional optical instruments (typically $\Delta z \approx 0.°01$). However, other values of $\Delta z$ can be relevant for simple sky scanners or all-sky cameras. Other modelling parameters are as follows: $z_0 = 73°$, $R(z, z_0) = 0.5$, $\Delta R = 0.02$.

## 4 DISCUSSION

The contribution of single scattering to the total radiance peaks at a small angular distance from the Sun (see Fig. 2). However, a small imperfection in reading radiance data near the solar disc can potentially increase the uncertainty of $\tau$ when the interpretation is based on equation (3). A separation distance of a few tens of degrees from the solar position on the sky is therefore preferred, which allows errors originating from inaccurate adjustment of positions along the circular arc to be minimized (see Figs 1–3). The theoretical error of $\tau$ depends on positioning inaccuracies $\Delta z$ and errors of the ratio $R(z, z_0)$; that is,

$$(\Delta \tau)^2 = \left(\frac{\partial \tau}{\partial R}\right)^2 \Delta R^2 + \left(\frac{\partial \tau}{\partial z}\right)^2 \Delta z^2 \ . \quad (10)$$

The solar zenith angle is assumed to be determined accurately from known Sun-position tools; thus, $\Delta z_0^2$ is considered to be much smaller than any of the above errors. Equation (10) can be also expressed as

$$\Delta \tau = \left|\frac{\partial R}{\partial \tau}\right|^{-1} \sqrt{(\Delta R)^2 + \left(\frac{\partial R}{\partial z}\right)^2 (\Delta z)^2} \ , \quad (11)$$

where partial derivatives can be obtained from equation (3):

$$\frac{\partial R}{\partial \tau} = -\frac{R}{\tau} - (\sec z - \sec z_0)\left(R + \frac{\cos^2 z_0}{\cos z - \cos z_0}\frac{1}{\tau}\right) \quad (12)$$

and

$$\frac{\partial R}{\partial z} = \frac{R \sin z}{\cos z - \cos z_0} - \frac{\sin z}{\cos^2 z}\left(R\tau + \frac{\cos^2 z_0}{\cos z - \cos z_0}\right) \ . \quad (13)$$

The relative importance of $|\partial R/\partial \tau|^{-1}$ and $|\partial R/\partial z|$ is analysed in the left-hand plot of Fig. 6. While the first term generally decreases as the direction of observation approaches the horizon (solid lines in the figure), the second term can peak near the position of the Sun (open and filled circles). This suggests that accurate sky scanner adjustment (or high image resolution) is especially important when the radiance reading is for sky elements near the Sun and $\tau$ is determined from equation (3).

Even if the positioning system is perfectly accurate (i.e. $\Delta z \approx 0$), $\Delta \tau$ still scales proportionally to the experimental errors of $R(z, z_0)$. Because $R(z, z_0) = L(z)/L(z_0)$, $\Delta R^2$ is a linear combination of quadratic forms for the errors of $L(z)$ and $L(z_0)$:

$$(\Delta R)^2 = \left[\left(\frac{\Delta L(z)}{L(z)}\right)^2 + \left(\frac{\Delta L(z_0)}{L(z_0)}\right)^2\right] R^2. \quad (14)$$

Assuming that the reading error is a constant value for an optical instrument (i.e. $\Delta L(z) \approx \Delta L(z_0) \equiv \Delta L$), equation (14) reduces to

$$\Delta R = \left|\frac{\Delta L}{L(z_0)}\right|\sqrt{1+R^2}, \quad (15)$$

taking into account the identity formula $L(z) = R(z, z_0)L(z_0)$. The value of $L(z_0)$ taken at the solar almucantar typically exceeds that of $L(z)$ when $z < z_0$. For instance, EKO Instruments scanners for luminance measurements (Suárez-García et al. 2021) have calibration errors of $(\Delta L)/L \approx 2$ per cent. For this value, $\Delta \tau$ is shown in the right-hand plot in Fig. 6 first for an ideally accurate positioning system ($\Delta z = 0$) and then for $\Delta z = 0.°1$ and $\Delta z = 0.°3$. The computational results demonstrate that radiance readings near the solar disc must be avoided when $\tau$ is to be retrieved from equation (3). Assuming that the angular distance from the solar disc does not exceed $30° - 40°$, the maximum error of the atmospheric optical depth is $\sim 0.03$. For the value of $\tau = 0.3$ we used in the numerical test, the relative retrieval error we could expect is smaller than 10 per cent (0.03/0.3). For the device we used, $(\Delta L)/L \approx 0.3$ per cent (Kocifaj et al. 2018), so $\Delta \tau$ is even smaller.

On the other hand, equation (4) provides a more stable solution for $g$ when observations are for small scattering angles. Owing to its potential uncertainty, the ratio of AOD/ROD can make the retrieval of $g$ inaccurate if neither $AOD \times P_A(\theta)$ nor $ROD \times P_R(\theta)$ has a dominant contribution to equation (5). In most situations, $P_A(\theta) \gg P_R(\theta)$ for $\theta < \approx 30°$, and thus the ratio of $P(\theta_1)/P(\theta_2)$ in equation (4) differs only slightly from its asymptotic value $P_A(\theta_1)/P_A(\theta_2)$. A choice of small scattering angles then efficiently suppresses retrieval uncertainties for $g$.

The error analysis indicates that amplitudes of $\Delta L$ larger than any typical fluctuations of the clear daytime sky radiance can significantly affect the success of the method developed here. However, such large amplitudes are typically associated with the presence of mist or may appear during cloud formation; that is, in situations in which astronomical observations are difficult or impossible to make. Because the method is exclusively designed for clear-sky conditions, the formulae derived in this paper are applicable each time astronomical observations are possible; in fact, this was the primary motivation for constructing the model.

The method is suitable for narrowband sky radiometry, where AOD and $g$ are spectral values, but correspondingly the broadband values can be obtained using, for example, RGB sky imaging. In that case, the photometric $k$-coefficients are related to the average over the corresponding passband's optical depth: $k \approx 2.5 \log_{10}(e) \tau$ magnitudes per airmass. So, this method could also predict $k$-values for a given night. The above formula provides an estimate for $k$, where $\tau$ is for the wavelength at the centre of the corresponding photometric filter. Of course, both the aerosol parameters and the solar flux are wavelength-dependent. An exact formula would include the spectrum integrated over a bandwidth rather than the use of a single-value approximation for $\tau$ centred on the bandwidth's nominal wavelength.

## 5 CONCLUSIONS

We have developed a method to retrieve the AOD and aerosol asymmetry parameter $g$ from clear-sky radiance taken during stable synoptic situations shortly before sunset or after sunrise. Such aerosol properties are of great importance in modelling NSB at the measuring site. Both these parameters are indispensable for predicting the sphere of influence of artificial light sources, while accepting that the state of the atmosphere is always momentary.

Optical depth is extracted from the ratio of radiance data recorded at different zenith angles, but at a constant small angular distance from the solar disc. A small separation distance from the Sun is preferred because the weighted contribution from single scattering to the total radiance reaches its peak values near the Sun's position. It is convenient to read the radiance first at the solar almucantar and then retrieve the corresponding value on the solar vertical line for an observational zenith angle $z < z_0$. The atmospheric optical thickness $\tau$ is determined by matching the theoretical and experimental ratios for the above two radiance values. The ROD varies only slightly with atmospheric pressure, so its contribution to the model is easy to compute theoretically. AOD is then obtained by subtracting ROD from $\tau$. The aerosol asymmetry parameter is determined from radiance data measured at different angular distances from the Sun, taking into account that $\tau$ is known. Small scattering angles are preferred in order to eliminate uncertainty in the AOD/ROD ratio.

The main benefit to the approach in this paper is that it does not require specialized (and expensive) equipment such as AERONET stations or light detection and ranging (LIDAR) equipment. Observatories might be able to extract the required imagery directly from their existing all-sky cloud cameras. We have shown that the theoretical retrieval error should not exceed 10 per cent for average atmospheric turbidity. We believe that the method is capable of improving the characterization of atmospheric conditions at astronomical observatory sites on nights when clear-sky conditions exist. A potential weakness of this approach is that conditions could change during the night. Measuring aerosol parameters at sunset does not obviate the need to also measure optical parameters (including $k$-coefficients) during the night, but it could help to inform observing plans if it was known at sunset that extinction was expected to be high or low.

## ACKNOWLEDGEMENTS

This work was supported by the Slovak Research and Development Agency under contract no. APVV-18-0014. Computational work was supported by the Slovak National Grant Agency VEGA (grant no. 2/0010/20).

## DATA AVAILABILITY STATEMENT

The simulation data underlying this article and radiance data gathered in the field experiment will be shared on reasonable request to the corresponding author.

This paper has been typeset from a TEX/LATEX file prepared by the author.